\documentclass[aps,prc,showpacs,twocolumn,superscriptaddress]{revtex4-1}
\usepackage{amsmath,amsfonts,amssymb,amsthm,float,mathastext, array, booktabs, tabularx, mathtools}
\usepackage{booktabs}
\usepackage{graphicx}
\graphicspath{{fig/}}
\usepackage{siunitx}
\usepackage{hyperref}
\hypersetup{hidelinks,colorlinks=false,breaklinks=true,urlcolor=ocre}
\usepackage{dcolumn}
\usepackage{bm}
\usepackage[english]{babel}
\usepackage[autostyle]{csquotes}
\usepackage{color}
\begin{document}
\preprint{APS/123-QED}

\title{Observation of annual modulation induced by $\gamma$ rays from ($\alpha$, $\gamma$) reactions at the Soudan Underground Laboratory}
\author{Ashok Tiwari}
 \affiliation{Department of Physics, The University of South Dakota, Vermillion, South Dakota 57069, USA}
 \author{C. Zhang}
 \affiliation{Department of Physics, The University of South Dakota, Vermillion, South Dakota 57069, USA}
\author{D.-M. Mei} 
 \email{Corresponding author.\\Email: Dongming.Mei@usd.edu}
 \affiliation{Department of Physics, The University of South Dakota, Vermillion, South Dakota 57069, USA}%
\affiliation{School of Physics and Optoelectronic, Yangtze University, Jingzhou 434023, China}
\author{P. Cushman}
\affiliation{School of Physics and Astronomy, University of Minnesota, Minneapolis, MN, 55455}
\date{\today}
\begin{abstract}
Annual modulation of $\gamma$ rays from ($\alpha$, $\gamma$) reactions in the Soudan Underground Lab has been observed using a 12-liter scintillation detector. This significant annual modulation, measured over 4 years, can mimic the signature for dark matter and can also generate potential background events for neutrinoless double-$\beta$ decay experiments. The measured annual modulation of the event rate from  ($\alpha$, $\gamma$) reactions is strongly correlated with the time-varying radon concentration observed independently in the Lab. The $\alpha$ flux from radon decay is simulated starting from the measured radon concentration, and the $\gamma$-ray flux is determined using the convolution of the $\alpha$ flux and the cross sections for ($\alpha$, $\gamma$) reactions. The calculated $\gamma$-ray flux is sufficient to generate the measured event rate that exhibits an annual modulation.            
\end{abstract}
\maketitle
\section{Introduction}
Observations starting from the 1930s~\cite{fzw} have led to the understanding that 80\% of the matter in the universe neither emits nor absorbs electromagnetic radiation~\cite{ghi, rjg}. A popular candidate for dark matter is the WIMP (Weakly Interacting Massive Particle)~\cite{jlf, mwg}, a fundamental particle with a mass of tens to hundreds of GeV and interactions at the weak scale.  In the past decade, a number of theories~\cite{res, cmh} have renewed interest in looking for light (MeV-scale) dark matter, giving rise to new proposals for experimental technologies that can detect extremely low energy depositions.   Both types of dark matter should present an annual modulation signal due to the relative motion of the Earth around the Sun~\cite{freese}, thus providing a critical signature which would confirm the dark matter hypothesis, when confronted with a significant measured excess.   While most background sources cannot mimic the details of the expected annual modulation~\cite{freese, dama, cogent2}, there are exceptions, such as muon-induced processes~\cite{jch} and radon concentration~\cite{ruddick, minos} in mine drifts.   With better understanding of these backgrounds, we can improve our  ability to detect and confirm dark matter signals, as well as improve the sensitivity of other rare-event experiments such as the search for neutrinoless double-$\beta$ decay. 

Backgrounds can be classified as either radiogenic or cosmogenic. Cosmogenic backgrounds come from cosmic rays, where the resulting muons produce high energy neutrons in the surrounding rock via the muon spallation process~\cite{muon}.  Radiogenic backgrounds are produced by natural radioactivity, such as uranium, thorium, and potassium decays.  In the radiogenic processes, $\gamma$ rays and neutrons are  generated by natural radioactivity decays, $\alpha$ interactions with surrounding materials, and the spontaneous fission decay of uranium and thorium in rocks. The fluxes of $\gamma$ rays and neutrons depend on the underground site and the elemental make-up in the surrounding materials.  Radon, a source of radiogenic background, represents a major threat to all rare-event experiments performed in underground labs. Though the level of radon can be mitigated by using proper ventilation and air circulation inside the underground laboratory, the residual radon concentration inside the air depends on the local geology (production) and experimental spaces (efficacy of mitigation).  Thus, radiogenic backgrounds are independent of the depth of the laboratory, whereas cosmogenic processes are related through their muon predecessors to the depth of the underground site~\cite{muon intensity}.

Several authors ~\cite{mei,heaton1} have calculated $\gamma$-ray and neutron yields based on ($\alpha$, $n$) reactions in different underground sites, where $\alpha$ particles are from natural radioactivity decays. However, little attention has been paid to ($\alpha$, $\gamma$) reactions. The decay chain of radon and thoron produces $\alpha$ particles with a few MeV of energy, which can interact with target elements and produce single $\gamma$ ray or cascades of $\gamma$ rays. The $\gamma$-ray yield from different targets can be calculated using the formula below~\cite{mei,feige,heaton2}: 
\begin{equation}
Y_{i} = \frac{N_{A}}{A_{i}} \sum_{j} \frac{R_{\alpha}(E_{j})}{S_{i}^m{(E_{j})}} \int_0^{E_{\alpha}} \frac{d\sigma_{i}{(E_{\alpha},E_{\gamma})}}{dE_{\alpha}}\;\mathrm{d}E_{\alpha}
\label{eq:gammayield}
\end{equation}
where, $A_{i}$ is atomic weight of $i^{th}$ element, $N_{A}$ is Avogadro's number, $S_{i}^m{(E)}$ is mass stopping power, and $\sigma_{i}{(E)}$ is the cross-section. $R_{\alpha}(E_{j})$ defines the production rate of the $\alpha$ particles from the radon decay chain with energies $E_{j}$. Since $\alpha$ particles lose energy continuously in materials, we require the integration over the total range of the energy deposition.

The Soudan mine is located at a 2100 meter water equivalent (m.w.e) i.e $\sim$710 meters underground. Several experiments, namely CoGeNT~\cite{cogent}, MINOS~\cite{minos} and CDMS~\cite{cdms} were performed within the Soudan mine. Since CoGeNT~\cite{cogent2} observed an annual modulation consistent with dark matter, it is important to investigate what background sources can reproduce this result. This paper describes the reactions of ($\alpha$,$\gamma$) for several targets, the fluxes of those $\gamma$ rays, and the radon concentration inside SUL, noting the correlation between our measured $\gamma$-ray induced annual modulation and the time-varying radon concentration in the lab.
 
\section{Detector and Calibration}
In order to measure backgrounds from radiogenic and cosmogenic processes, a 12-liter liquid scintillation detector~\cite{cmei} was installed at SUL. The detector is a cylindrical vessel which has dimensions of 1 m in length and 13 cm in diameter~\cite{zhang}. The detector volume was filled with  EJ301 liquid scintillator.  EJ301 is specially designed for neutron-$\gamma$ pulse shape discrimination~\cite{knoll}, with a H/C ratio 1.212 and density 0.874 $g/cm^3$.  Two photomultiplier tubes (PMTs), one at each end of the cylinder were used to collect the scintillation light produced inside. The light yield from EJ301 is 78$\%$  of anthracene, with photons of maximum wavelength 425 nm. This corresponds to the most sensitive region of the 5" R4144 Hamamatsu PMTs used. Further details of the detector and its readout system can be found in here~\cite{cmei}.  The detector operated for more than 4 years at SUL collecting $\gamma$-ray events generated inside the laboratory from ($\alpha$, $\gamma$) reactions due to both radon and thoron decays. The following criteria were used to select the $\gamma$-ray events:
\begin{enumerate}
\item {the events should not saturate the ADC channels;}
\item {both PMTs must trigger; and}
\item {their time coincidence must be within a 30 ns window defined by the two largest samples in the pulse collected by the PMTs.}
\end{enumerate}

The detector was calibrated using $^{22}$Na and AmBe $\gamma$ ray sources, as well as with a simple muon telescope. The AmBe source emits 4.44 MeV $\gamma$-rays and was used to calibrate the energy scale, which was verified by the minimum-ionizing muon tracks. The $^{22}$Na source gives two distinct $\gamma$ ray lines at 0.511 MeV and 1.275 MeV.  The $\gamma$ rays passed through a lead collimator and were used to calibrate the position scale by coincidence measurements along the length of the detector, every 5 cm (20 sections in all).  

\begin{figure} [H]
\centering
\includegraphics[scale= 0.45]{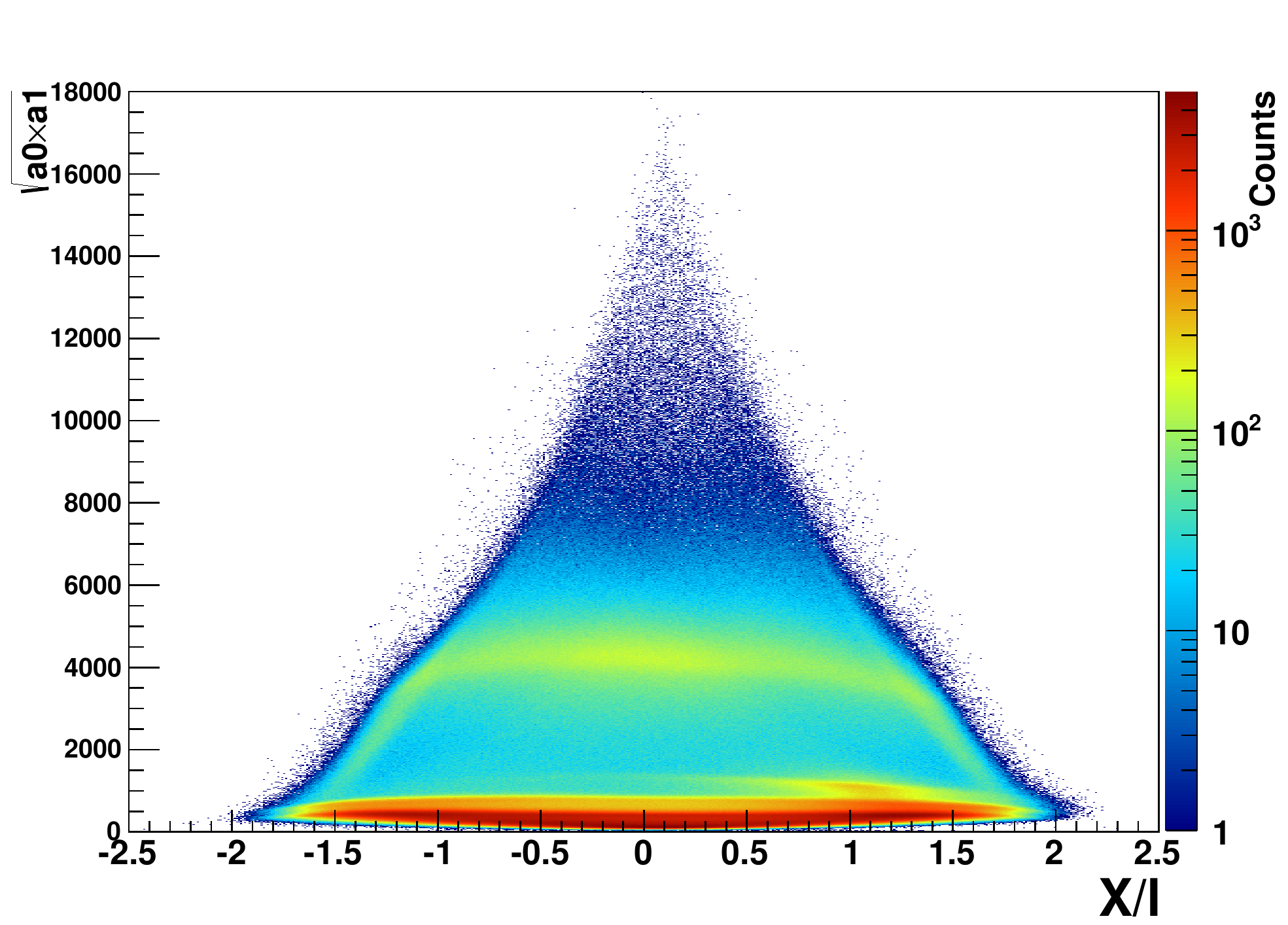}
\caption {(Color online). Plot of $\sqrt[]{a_{0}\times a_{1}}$ vs $X/l$ \cite{mei}, \cite{cmei}. Where, $ \sqrt{a_{0}\times a_{1}}$ represents the total energy deposited in the detector, $a_{0}$ and $a_{1}$ are the charge collected at the two ends of the detector and $X/l$ describes the relative position of the particle in the detector, $l$ is the attenuation length and $X$ is the distance between the mid point of the detector and the point where a high energy event is deposited.}
\label{fig:calibration}
\end{figure}

Fig.~\ref{fig:calibration} summarizes the results from the multiple calibration procedures described above and requires some explanation. Its overall triangular shape is defined by the fact that events near the ends of the cylindrical detector saturate one of the ADC channels rendering the data unusable.  This happens more frequently, the higher the energy of the event.  The faint, lower yellow band is due to the AmBe source and the higher yellow energy band shows the response to minimum ionizing muons ($\sim$20 MeV deposited in this detector).  The horizontal nature of the yellow bands demonstrate that the detected energy is independent of its position along the tube, where $X/l$=0 represents the mid point of the detector. The red in the lower part of the graph is due to a local concentration of radioactivity.
\section{Annual modulation of the $\gamma$-ray flux}
\begin{figure} [H]
\centering
\includegraphics[scale=0.45]{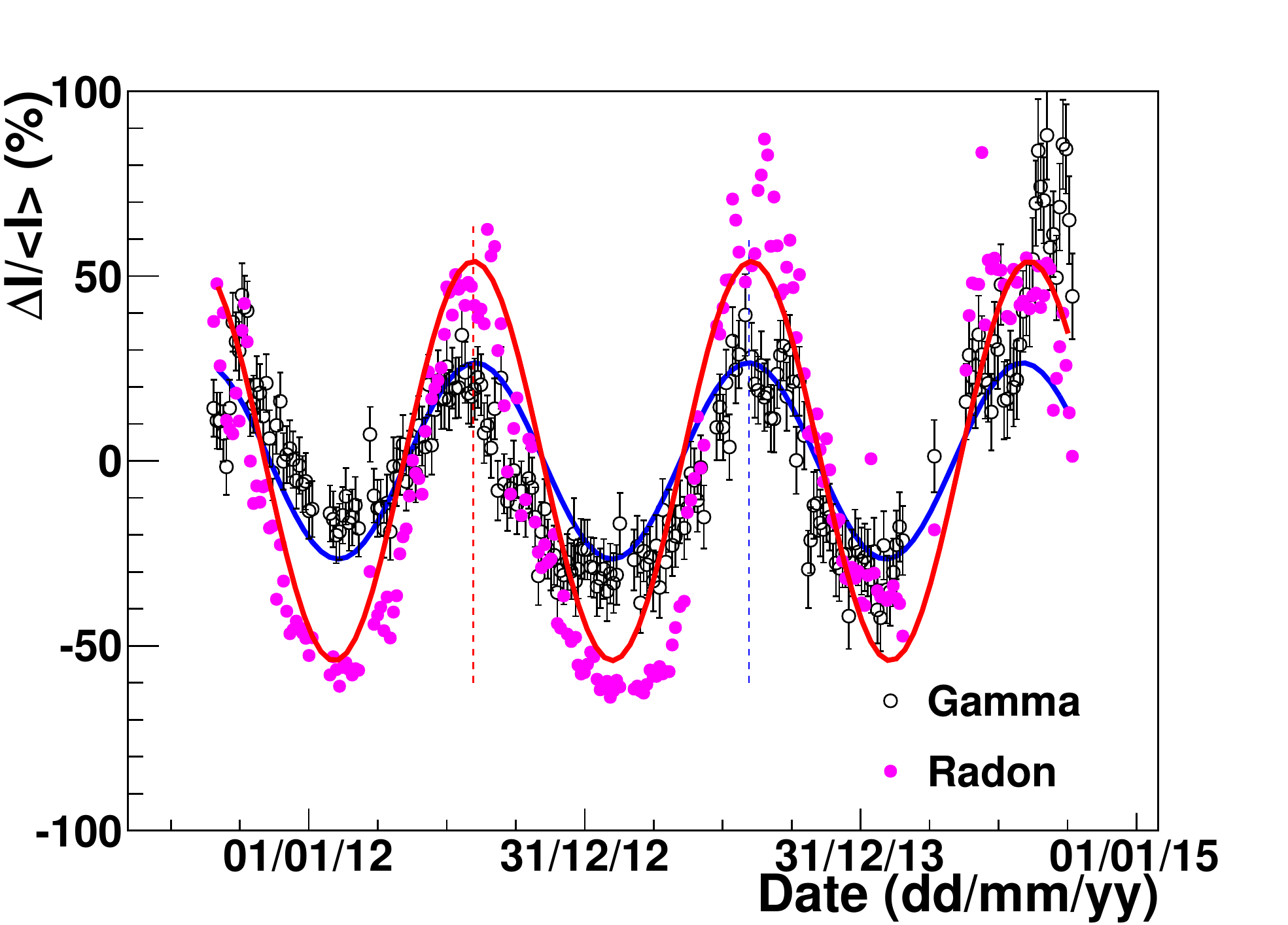}
\caption  {(Color online). Annual modulation of $\gamma$-flux in the Soudan Underground Lab. The x-axis is calendar time from August 28, 2011 to October 8, 2014 (covering $\sim$4 years), while the y-axis represents the percent change in $\gamma$ ray intensity measured by the scintillator detector. The blue line is the best fit sinusoidal modulation to the black circles which represent selected $\gamma$ events. Also plotted is the percent change in radon concentration for the same period of time, where the data (magenta dots) was collected by MINOS and the red line is a best fit using Eq.~\ref{mod}.} 
\label{fig:gamma_mod}
\end{figure}

A sinusoidal time dependence of the $\gamma$-ray fluxes in the energy region between 4 - 10 MeV was observed, as shown in Fig.~\ref{fig:gamma_mod}. The data from selected $\gamma$ events measured in the scintillation detector are the black circles, binned every 4.6 days and plotted as the ratio $\Delta$$I/I$ ($\%$) which  is a measure of the amplitude modulation. The higher $\gamma$ flux near the end of 2014 is not understood, but is included in both fit and correlation, increasing our systematic uncertainty.

The modulated $\gamma$-ray flux was fit to a simple equation:  
\begin{equation}
\label{mod}
I = I_{0} + \Delta I = I_{0} + \delta I cos(\frac{2\pi}{T}(t-t_{0})),
\end{equation}
where $I_{0}$ is the un-modulated $\gamma$-ray flux and $t_{0}$ is the measured phase of modulation. The phase t$_{0}$ is defined as the day at which the signal is at a maximum. The fit is given by the blue line in Fig.~\ref{fig:gamma_mod}. The modulation amplitude was found to be (1.73 $\pm$ 0.45)$\times$$10^{-7}$ $cm^{-2} s^{-1}$, where the uncertainty is purely statistical.  This modulation corresponds to 26.5$\%$ of the total average intensity, $I_{0}$ = (5.76 $\pm$ 0.98) $\times$$10^{-7}$ $cm^{-2} s^{-1}$.  We found a period, $T$, of 362.4$\pm$1.8 days with a phase that peaks on August 06. 

\section{Correlation with radon concentration}
\begin{figure} [H]
\centering
\includegraphics[scale= 0.23]{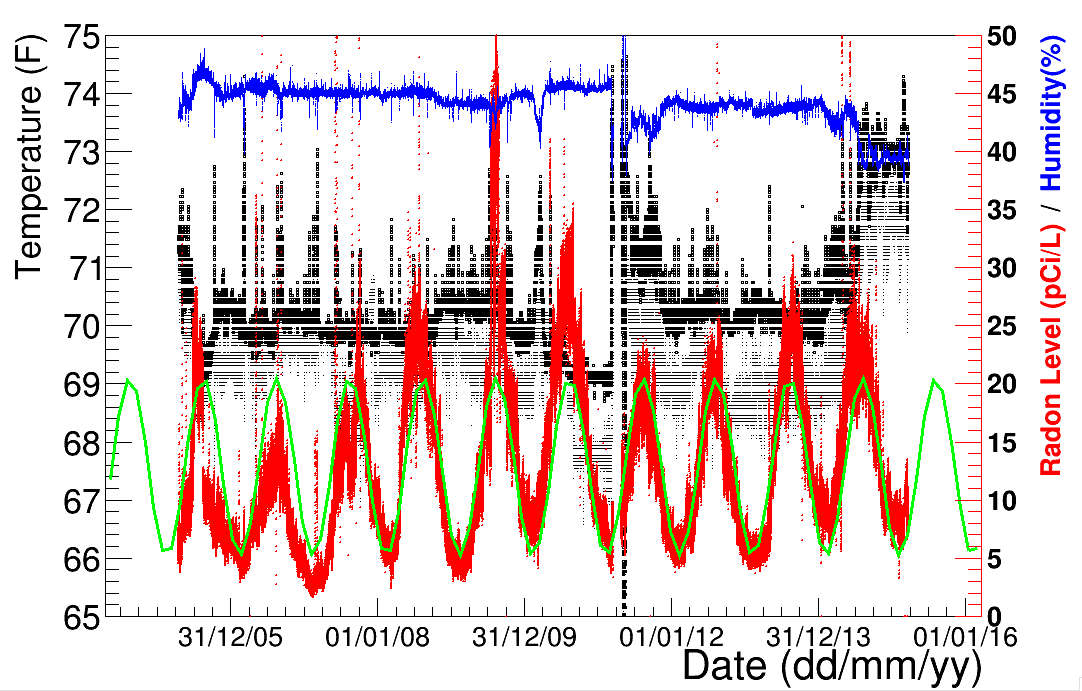}
\caption  {(Color online). Radon modulation data (red) from the MINOS collaboration~\cite{minos} plotted against temperature (black) and humidity (blue).  The green line is an Eq.~\ref{mod} fit to the radon concentration seasonal variation over ten years, giving a slightly different period and phase than the four year fit.} 
\label{fig:radonvariation}
\end{figure}

The correlation between the measured $\gamma$ flux and the radon concentration inside the laboratory was examined over 4 years of data.  It is well-known that radon levels fluctuate due to seasonal air flow reversals within the larger Soudan mine complex~\cite{radon1}.  According to the MINOS Collaboration~\cite{minos}, the radon level inside the laboratory varies from a winter low of 5.0 pCi/Liter to a summer high of 
25.0 pCi/Liter, as shown in Fig.~\ref{fig:radonvariation}. The measured $\gamma$ flux strongly correlates with the radon variation inside the laboratory, as shown in Fig.~\ref{fig:gamma_mod}, where the percent change in the radon concentration data (magenta) and its best fit (red) are plotted for the same time period. Note that the modulation amplitude of the radon concentration is determined to be 54\% of its average concentration of 15 pCi/Liter and the period is 368.0$\pm$0.12 days with a phase that peaks on August 06, which is exactly same as the measured $\gamma$ rays.

Since the observed events are in the energy region of 4 - 10 MeV, we attribute these events to Compton scatters as well as full absorption of $\gamma$ rays.  As the observed $\gamma$-ray flux is rather large, at the level of $\sim$5.8$\times$10$^{-7}$cm$^{-2}$s$^{-1}$, the modulation cannot be coming from the much smaller muon-induced high energy $\gamma$ rays, which is  $\sim$10$^{-9}$cm$^{-2}$s$^{-1}$~\cite{cmei}. 

As a check, the radon concentration shows no correlation with temperature and humidity inside the laboratory, as shown in Fig.~\ref{fig:radonvariation} over a much longer period of ten years (with 5 minutes interval) of tracking by the Soudan Lab.  The solid  green line represents the best fit using Eq.\ref{mod}.  A possible instrumentation error up to 5\%  may exist with the radon  concentration data measured with RAD 7 detectors~\cite{rad7}

\begin{figure} [H]
\centering
\includegraphics[scale= 0.45]{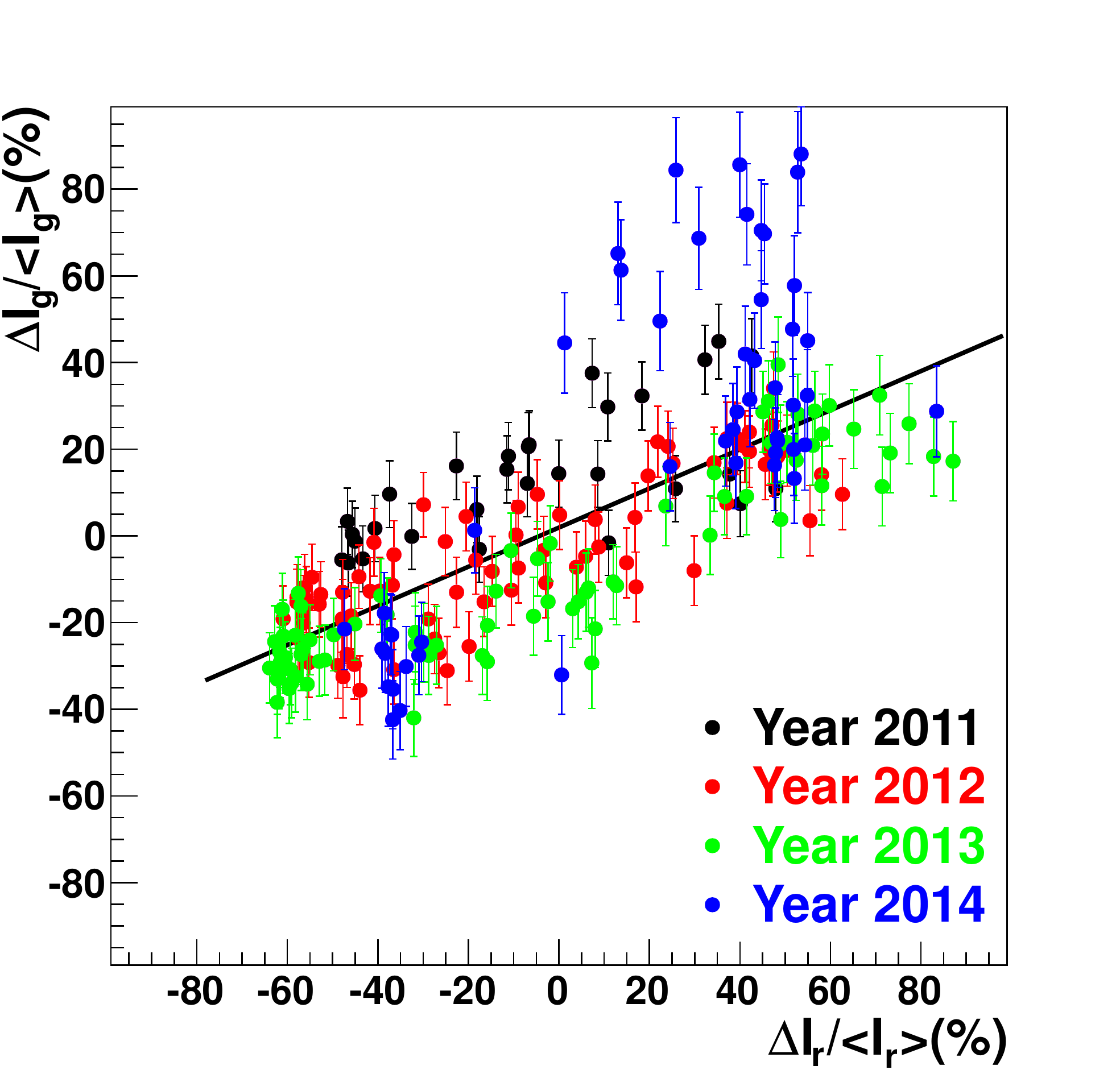}
\caption  {Correlation of the percentage variation of $\gamma$-ray flux (y-axis) versus the percentage variation of radon rates (x-axis) at SUL. The correlation coefficient between $\gamma$ rays and radon is found to be (73.9$\pm$ 21.7)$\%$.} 
\label{fig:corplot}
\end{figure}

Using the measured $\gamma$-ray modulation amplitudes and the radon modulation amplitudes from Fig.~\ref{fig:gamma_mod}, a correlation plot (Fig.~\ref{fig:corplot}) can be generated. 
From Fig.~\ref{fig:gamma_mod}, it is clear that radon and $\gamma$-ray modulation amplitudes are proportional to each other. Therefore, we can write:
\begin{equation}
R^{a} = m G^{a},
\end{equation}
where, $m$ is the proportionality constant which gives the slope of Fig.~\ref{fig:corplot} and $R^{a}$ and $G^{a}$ are the measure of radon and $\gamma$-yield amplitudes respectively from Fig.~\ref{fig:gamma_mod}. The degree of correlation has been evaluated using Pearson's correlation coefficient method, calculated as:
\begin{equation}
\rho = \frac{\sum\limits_{n}(G^{a}-\overline{{G^{a}}}) (R^{a}-\overline{{R^{a}}})}{\sqrt[]{\sum\limits_{n}{(G^{a}-\overline{{G^{a}}})^2}}\sqrt[]{\sum\limits_{n}{(R^{a}-\overline{{R^{a}}})^2}}},
\end{equation}
where, $\overline{{G^{a}}}$ and  $\overline{{R^{a}}}$ are average amplitudes of $\gamma$-yield and radon level. The slope of the line in Fig.~\ref{fig:corplot} is 0.451$\pm$0.013 and the correlation coefficient is 0.739$\pm$0.217, where the uncertainty is pure statistical error. 

A measure of the systematic error associated with the correlation coefficient can be evaluated by comparing the data from each individual year. The center value of the correlation coefficient from each year  is subtracted from the 4-yr value reported above.  The average of this difference over the four years is 0.145, so the measured correlation coefficient can be quoted as 0.739$\pm$0.217(stat)$\pm$0.145(sys).
  
\section{Modeling $\gamma$-ray fluxes at SUL}
The $\gamma$-ray yield is given by Eq.~\ref{eq:gammayield}
and thus depends on the incident energy of the initial $\alpha$ particle, the stopping power of the $\alpha$ within the target and the differential reaction cross section of the ($\alpha$, $\gamma$) reaction.  We now go through each of the terms used in this equation.  The stopping power (sum of electronic and nuclear) was obtained from the ASTAR database ~\cite{astar}. 

The energy of the $\alpha$ is determined by their production process in SUL.  The radiogenic backgrounds at Soudan are dominated by radioactive elements in the rock, such as $^{238}$U, $^{232}$Th and $^{40}$K, having half lives 4.468, 14.05 and 1.3 billion years respectively ~\cite{data}. The decay chains of $^{238}$U and $^{232}$Th produce radon and thoron, which generate $\alpha$ particles with energy in the range of few MeV as shown in Table~\ref{table1}.  
\begin{table}[h!]
\centering
\begin{tabular}{|c|c|c|c|}
\hline 
Decay mode (radon) & Alpha energy (MeV) \\ 
\hline
$^{222}_{86}Rn$ $\rightarrow$ $^{218}_{84}Po$ + $\alpha$ & 5.590 \\ 
\hline 
$^{218}_{84}Po$ $\rightarrow$ $^{214}_{82}Pb$ + $\alpha$ & 6.115 \\ 
\hline 
$^{214}_{84}Po$ $\rightarrow$ $^{210}_{82}Pb$ + $\alpha$ & 7.833 \\ 
\hline
$^{210}_{84}Po$ $\rightarrow$ $^{206}_{82}Pb$ + $\alpha$ & 5.407 \\
\hline
\hline
Decay mode (thoron) & Alpha energy (MeV) \\ 
\hline
$^{220}_{86}Rn$ $\rightarrow$ $^{216}_{84}Po$ + $\alpha$ & 6.288 \\ 
\hline 
$^{216}_{84}Po$ $\rightarrow$ $^{212}_{82}Pb$ + $\alpha$ & 6.778 \\ 
\hline 
$^{212}_{83}Bi$ $\rightarrow$ $^{206}_{81}Tl$ + $\alpha$ & 6.090 \\ 
\hline
$^{212}_{84}Po$ $\rightarrow$ $^{208}_{82}Pb$ + $\alpha$ & 8.784 \\
\hline
\end{tabular}
\caption{Energy of alpha particles generated from radon ($^{222}_{86}Rn$) and thoron ($^{220}_{86}Rn$) decay chains \cite{isotop1}:}
\label{table1}
\end{table}

The term $R_{\alpha}(E_{j})$ in Eq.~\ref{eq:gammayield}, accounts for the production rate of $\alpha$ particles in the cavern. From Fig.~\ref{fig:radonvariation}, the average radon concentration in the cavern is calculated to be 15 p Ci/L, or 555 $Bq/m^{3}$. 
Since the thermal velocity of radon and its air-borne daughters is about 1.83$\times$10$^{4}$ cm/s, this provides a flux of particles ($\sim$ 10.16 $cm^{-2}s^{-1}$) that can intersect the target surface. The average range of $\alpha$ particles in air is approximately 3.2 cm. According to the Bragg-Kleeman rule \cite{bragg-kleeman}, the average range of  $\alpha$ particles in aluminum, silicon, and oxygen is 0.002 cm, 0.0023 cm and 3.07 cm respectively. Therefore, if $\alpha$ particles are produced by radon in the air, they cannot contribute to ($\alpha$,$\gamma$) reactions in any significant way and are neglected.   

Thus, there remain two ways to generate the $\alpha$ particles which will eventually produce a flux of $\gamma$ rays with energy in the range of 4 - 10 MeV. One is radon plate-out, where radon daughters adhere to the surface of the target, after which $\alpha$ particles are produced through subsequent decays. Secondly, since radon is a gas with a half life of 3.82 days, it can first diffuse into the target and then decay to generate $\alpha$ particles.  
 
The diffusion length in the aluminum walls of the detector can be written using the Fick's law~\cite{fick, ishimori} as: 
\begin{equation}
L=\sqrt[]{\frac{D}{\lambda}}
\label{ficks}
\end{equation}
where D is the diffusion coefficient for radon in aluminum in units of cm$^{-2}$s$^{-1}$ and $\lambda$ is the decay constant of radon in units of seconds. The decay constant of radon is $\lambda$=0.693/t$_{1/2}$ $\sim$2.31$\times$10$^{-6}$ s$^{-1}$. 

F. Mamedov et al.~\cite{fick} used aluminum foil to calculate the diffusion coefficient of radon in aluminum to obtain a diffusion coefficient of $\sim$5.1$\times$10$^{-11}$cm$^{2}$s$^{-1}$, which gives a diffusion length of $\sim$0.05 mm. Thus, the effective interaction zone in which ($\alpha$,$\gamma$) reactions occur is the combination of radon diffusion length in aluminum and the average range of $\alpha$ particles in aluminum. This combination is about 0.007 mm.  Similar effective interaction zones can be obtained for silicon and oxygen. 
 
Note that the thickness of our detector wall is $\sim$4 mm, which is much larger than than the radon diffusion length of 0.05 mm. Therefore, the chance of radon particles themselves entering the detector liquid volume via diffusion is close to zero.  The radon emanation measured from an aluminum metal plate~\cite{radon-emanation} is $<$0.5 per squared meter per hour, or $\sim$1.4$\times$10$^{-8}$cm$^{-2}$s$^{-1}$. Thus, this process cannot be responsible for the observed annual modulation signal.

\begin{figure}[H]
\centering
\includegraphics[scale= 0.45]{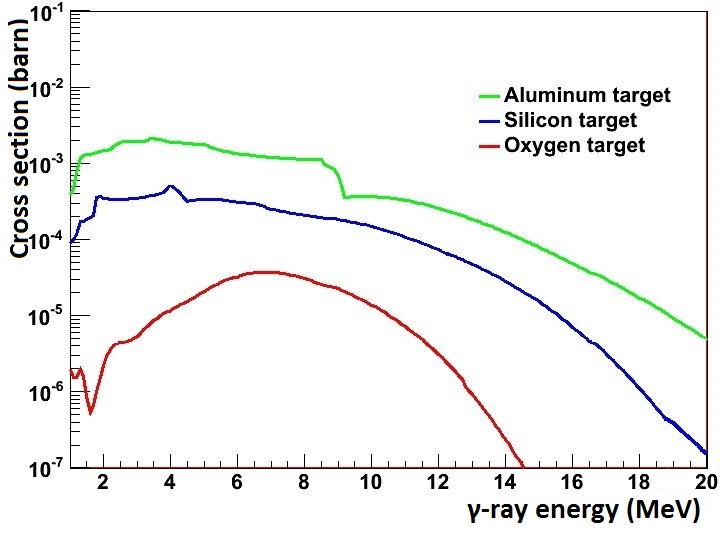}
\caption{(Color online). Calculation of cross section of the ($\alpha$, $\gamma$) reaction, when incident $\alpha$ particle has an energy 4 MeV for different targets, using TALYS code \cite{talys}. The green (upper) line is the cross section plot for the aluminum target, the blue (middle) line is for silicon and the red (lower) line is for oxygen target.} 
\label{fig:cross_sec}
\end{figure}

The rock composition at SUL is mainly Ely greenstone, which is composed of silicon, oxygen, aluminum, iron, calcium, and magnesium ~\cite{ruddick, rock}.  This rock composition is typical of basalt with a density $\rho$ = (2.75 - 2.95) $g/cm^2$.  In the calculation of the $\gamma$ ray flux from different targets, we choose the three most abundant of these elements: oxygen ($\sim$45\%), silicon ($\sim$24\%), and aluminum ($\sim$8\%). The excited states of these nuclei have $\gamma$ rays with energy greater than 6 MeV. Since $\gamma$ rays will mostly undergo Compton scattering in the detector, the energy deposition from those high-energy gamma rays can be in the energy region of 4-10 MeV. The cross section term in Eq.~\ref{eq:gammayield} was determined using the TALYS nuclear reaction modeling method~\cite{talys}.  Reaction cross sections were calculated for $\alpha$ particles which interact with the three different target elements considered. Fig.~\ref{fig:cross_sec} illustrates that the resulting cross section is highly energy-dependent. 

\begin{figure}[H]
\centering
\includegraphics[scale= 0.23]{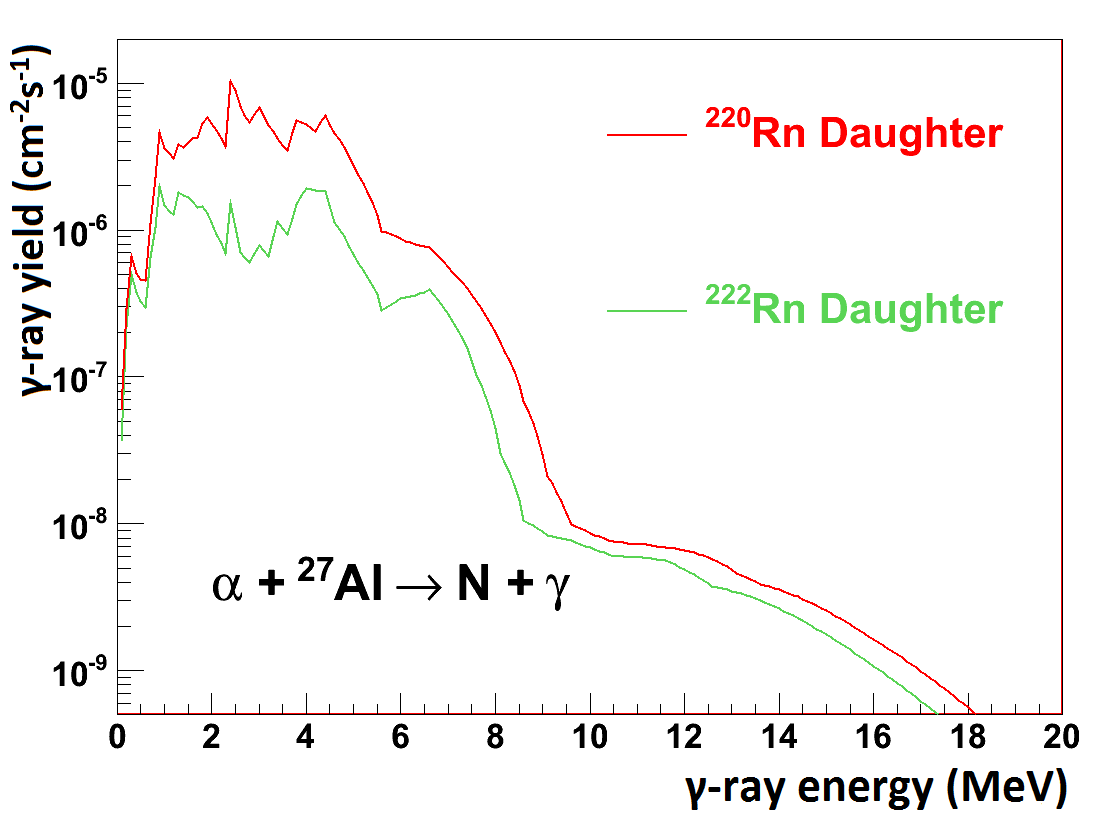}
\caption{(Color online). The $\gamma$-ray flux from the ($\alpha$, $\gamma$) reaction on the aluminum walls of the detector. The red (upper) line is the $\gamma$ yield due to thoron daughter and light green (lower) line is the $\gamma$ yield due to radon daughter.  The higher $\gamma$ yield in aluminum in the region from 2 - 8 MeV is not surprising since the reaction cross section is largest at low energy.} 
\label{fig:alu_gamma_yield}
\end{figure}

It is now possible to calculate the $\gamma$-ray yield from the most common isotopes of our three elements.  Although there are 22 isotopes of aluminum~\cite{isotop}, $^{27}Al$ is the only stable isotope of aluminum occurring naturally. $^{27}Al$ has a natural abundance of approximately 99.9$\%$.  The contribution from the $Al_{2}O_{3}$ rock component was also calculated, but contributes very little. Figure~\ref{fig:alu_gamma_yield} shows the calculated $\gamma$ ray-ray flux for the aluminum target. 

\begin{figure}[H]
\centering
\includegraphics[scale= 0.23]{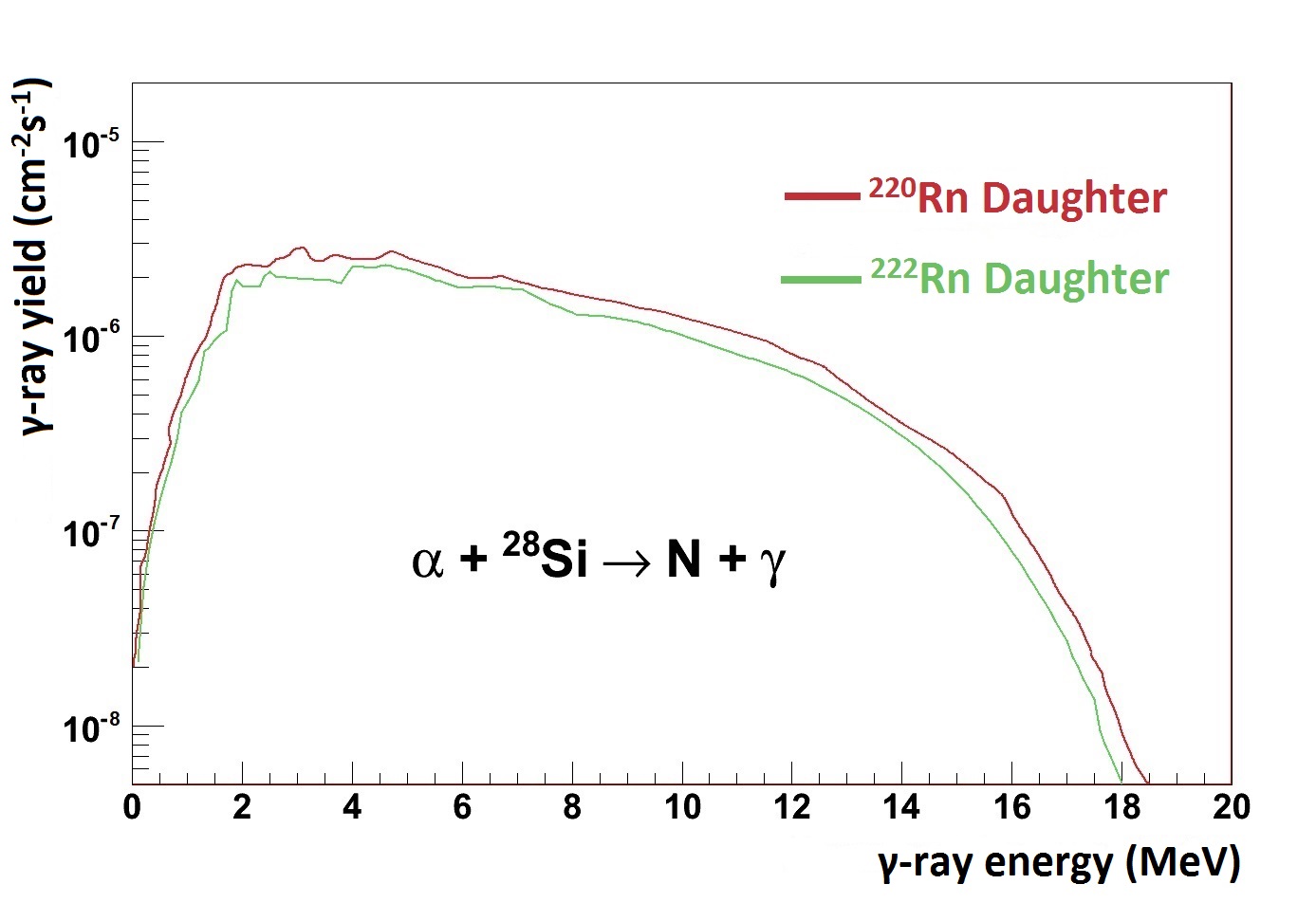}
\caption {(Color online). The $\gamma$-ray flux from ($\alpha$, $\gamma$) reaction in silicon (rock component). The red (upper) line is the $\gamma$ yield due to thoron daughter and light green (lower) line is the $\gamma$ yield due to radon daughter.The $\gamma$-ray yields depend strongly on the silicon present in rock cavities inside the mine. With increasing $\gamma$-ray energies, the flux gradually decreases.} 
\label{fig:Sil_gamma_yield}
\end{figure}

Silicon has 24 isotopes, of which three are stable: $^{28}Si$, $^{29}Si$, and $^{30}Si$~\cite{isotop1}. The natural abundance of isotope $^{28}Si$ is 92.22$\%$, $^{29}Si$ is 4.68$\%$ and $^{30}Si$ is 3.09$\%$. Therefore, $^{28}Si$ is used for the  calculation. Figure~\ref{fig:Sil_gamma_yield} displays the calculated $\gamma$-ray for the silicon target. 

\begin{figure} [H]
\centering
\includegraphics[scale= 0.22]{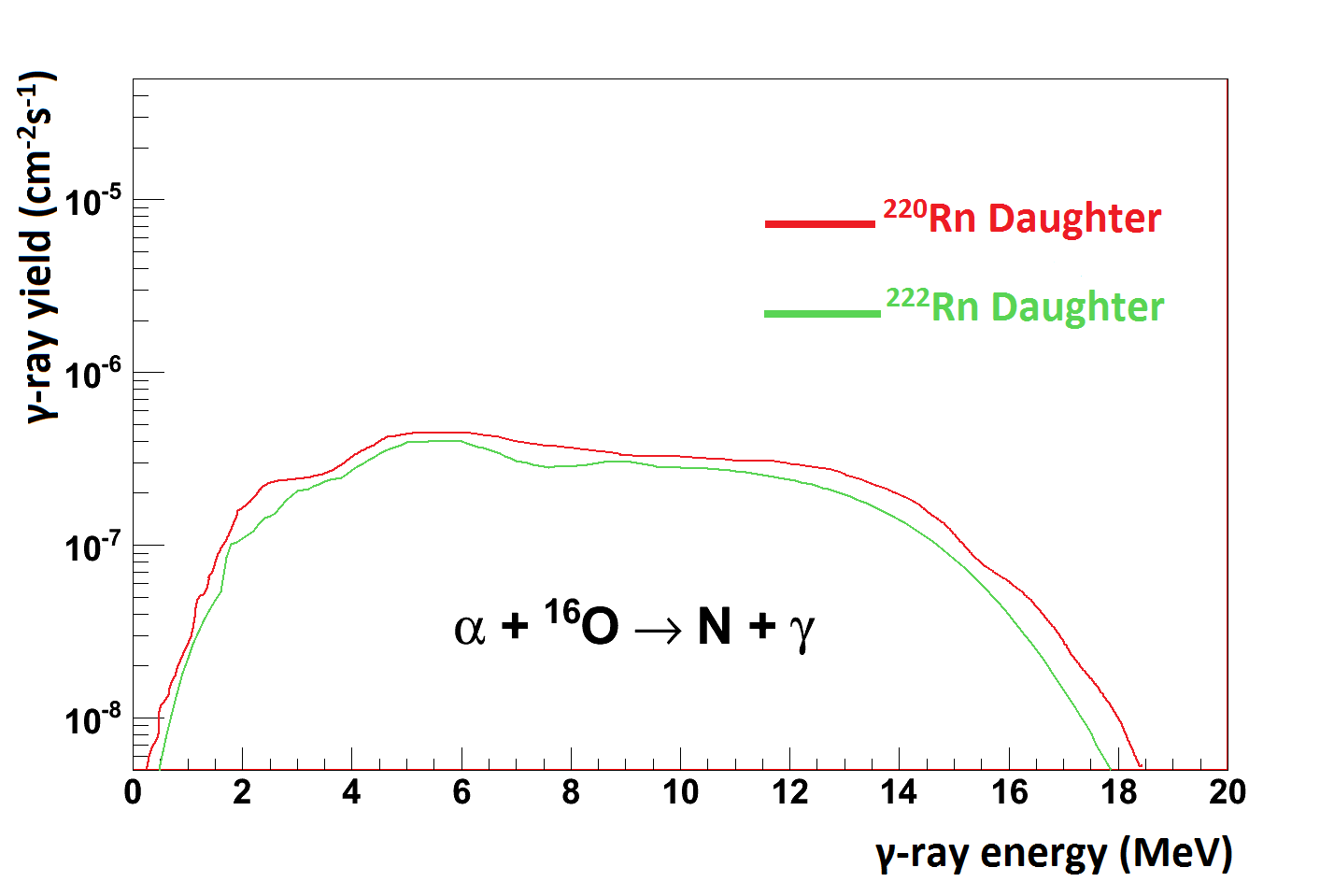}
\caption {(Color online). The $\gamma$-ray flux from ($\alpha$, $\gamma$) reaction with oxygen (air and rock component). The red (upper) line is the $\gamma$ yield due to thoron daughter and light green (lower) line is the $\gamma$ yield due to radon daughter. } 
\label{fig:oxy_gamma_yield}
\end{figure}

Oxygen isotopes are present everywhere inside the mine, in both the air and in the rock. There are 17 isotopes of oxygen, of which three are stable \cite{isotop2}: $^{16}O$, $^{17}O$ and $^{18}O$. Since $^{16}O$ has a natural abundance of 99.76$\%$, this is the isotope considered in the calculation. Since oxygen exists in air and in rock, the $\gamma$-ray yields were calculated separately and added together afterwards. Figure~\ref{fig:oxy_gamma_yield} exhibits the calculated $\gamma$-ray flux for the oxygen target. 

Figures~\ref{fig:alu_gamma_yield},~\ref{fig:Sil_gamma_yield},~\ref{fig:oxy_gamma_yield} are the plots of $\gamma$-ray flux in units of $cm^{-2}s^{-1}$ versus $\gamma$-ray energy in units of $MeV$ for different energy domains. The $\gamma$-ray fluxes are calculated for both radon and thoron decays since they yield $\alpha$ particles with slightly different energies. In all cases, the $\gamma$-ray yield falls off with energy above several MeV. 

The $\gamma$-ray flux calculated for the three different reaction channels is summarized in Table~\ref{table3}. These fluxes are integrated $\gamma$ fluxes in the energy region (4-10) MeV. It is clear that the aluminum and silicon targets yield a higher $\gamma$-ray flux than the oxygen.
\begin{table}[H]
\centering
\begin{tabular}{|m{1.8cm}|m{1.8cm}|m{1.8cm}|m{1.8cm}|}
\hline
	\toprule 
 	& Aluminum & Silicon & Oxygen \\
    \midrule
	\hline
 	$^{222}$Rn & $6.44\times 10^{-6}$ & $1.22\times10^{-6}$ & $3.54\times10^{-6}$\\ [1ex]
	\hline 
	$^{220}$Rn & $2.39\times 10^{-5}$ & $9.12\times10^{-5}$ & $5.18\times10^{-6}$\\ 
	\bottomrule
	\hline
\end{tabular}
\caption{Calculated values of integrated $\gamma$-ray fluxes for different reaction channels, (in units of $cm^{-2}s^{-1})$ for three different target elements: aluminum, silicon and oxygen. }
\label{table3}
\end{table}

\section{Conclusion}
\label{s:conc}
Using a liquid scintillation detector, $\gamma$-ray fluxes were measured in the energy region between 4 - 10 MeV in the Soudan mine and found to have an annual modulation similar to that expected from WIMP dark matter. The overall $\gamma$-ray rates are similar to measurements made by the NEMO collaboration at the Frejus underground laboratory (LSM)~\cite{frejus}, as shown in Table~\ref{table4}.
\begin{table}[H]
\centering
  \begin{tabular}{| >{\centering\arraybackslash}m{3cm} | >{\centering\arraybackslash}m{3cm}| >{\centering\arraybackslash}m{2cm} |}
  \hline
    \toprule
    Energy range (MeV) & SUL (this paper) & LSM (Ref.~\cite{frejus}) \\
    \hline
    \midrule
    4.0 - 6.0 & 1.20$\pm$0.36 & 3.8  \\ 
    6.0 - 7.0 & 5.60$\pm$1.68 & 1.5  \\
    7.0 - 8.0 & 7.80$\pm$2.34 & 1.6  \\
    8.0 - 9.0 & 0.010$\pm$0.003 & 0.07 \\ 
    9.0 - 10.0 & 0.04$\pm$0.01 & 0.05 \\
    \hline
    \bottomrule
  \end{tabular}
 \caption{The $\gamma$-ray fluxes in the Soudan Underground facility. The unit of $\gamma$ yield is $10^{-6}$$ {cm^{-2}s^{-1}}$. Results from similar measurements made at the Frejus Laboratory are also shown in the table. }.
    \label{table4}
\end{table}
The observed annual modulation of ($\alpha$, $\gamma$) induced events are positively correlated with the radon concentration. The $\gamma$ yields from different reaction channels show modulation patterns with amplitude (1.73 $\pm$ 0.45(sta))$\times$$10^{-7}$ $cm^{-2} s^{-1}$ with a statistical uncertainty on the observed modulation amplitude of $\sim$30$\%$. Since the $\gamma$ fluxes are larger than the measured amplitude modulation, $\gamma$ rays generated from different targets can separately exhibit annual modulation patterns.

Errors quoted in the $\gamma$-ray flux calculation come only from uncertainties in the calculation of the $\alpha$ particle production rate and cross sections. Without knowing the exact air circulation rate in the laboratory, it is impossible to determine how many radon daughters will be deposited in the detector after radon decay, thus contributing a possible source of systematic uncertainty to our result. 
 
In Table~\ref{com}, the annual modulation fit parameters from our scintillator detector are  compared to those expected from dark matter assuming a standard halo~\cite{kat}, as well as to the CoGeNT fit. While the amplitude and period of all three are similar, the phases differ significantly. CoGeNT finds an earlier maximum of mid-April, while this work finds an early August maximum.  The standard halo model for WIMPs predicts a maximum in early June. In this case, it is unlikely that the radon-induced ($\alpha$, $\gamma$) reaction is responsible for the signal observed by CoGeNT.  However, it is also clear that these reactions can easily create signals as large as the effect due to dark matter and thus need to be properly accounted for in any fit to a putative signal.  
\begin{table}[H]
\centering
\begin{tabular}{m{1.8cm}| m{2.0cm} |  m{1.9cm}| m{2.0cm} }
\hline
    \toprule 
 	Fit&Amplitude  ($\delta$I, \%)~ &Period \hspace{0.8cm} (T, days)&Phase ~\hspace{1.0cm}(t$_{0}$, days) \\
    \midrule
    \hline
    \vspace{0.1cm}
    WIMP Model~\cite{kat} & $\mathcal{O}$(1-10)&$\sim$365.25& $\sim$150\\	\hline
   \vspace{0.1cm}
 	CoGeNT & 16.6$\pm$3.8& 347$\pm$29& 115$\pm$12\\ 
	\hline 
    \vspace{0.1cm}	
    This work & 26.5$\pm$7.8 & 362.4$\pm$1.8& 218$\pm$1.1\\ 
    \hline
	\bottomrule
	\hline
\end{tabular}
\caption{Annual modulation fit parameters for CoGeNT compared to theory and to this work.}
\label{com}
\end{table}
In conclusion, this study describes a type of background that could be important for experiments related to the direct search for dark matter and neutrinoless double-$\beta$ decay. The significant annual modulation induced by $\gamma$ rays from ($\alpha$, $\gamma$) reactions can mimic the signature for dark matter and generate potential background events for neutrinoless double-$\beta$ decay experiments, since 4 - 10 MeV $\gamma$ rays are capable of penetrating an outer shield to undergo Compton scattering in an inner shield or undergo a photo-nuclear reaction which generates neutrons in an inner shield.  The $\gamma$ flux in the 4- 10 MeV range depends on the surrounding materials of the laboratory and rock cavities inside the mine, while energies above 10 MeV depend on the depth of the laboratory. 

 \section*{Acknowledgments}
The authors would like to thank the Soudan Underground Laboratory staff and technical team for their assistance in running the experiment. The authors appreciate the wonderful suggestions and comments from Jing Liu. This work was supported in part by NSF PHY-0919278, NSF PHY-1242640, NSF OIA 1434142, DOE grant DE-FG02-10ER46709, and a governor's research center supported by the state of South Dakota.

\end{document}